\documentclass[twocolumn,aps,prb,showpacs,superscriptaddress]{revtex4}
\usepackage{amsmath}
\usepackage{epsfig}
\usepackage{epsf}
\usepackage{array}
\usepackage{color}
\usepackage{graphicx}
\usepackage{epstopdf}

\begin{document}
\bibliographystyle{apsrev}

\title{Pure and Random-Field Quantum Criticality in the Dipolar Ising Model: Theory of $Mn_{12}$ acetates}
\author{ A. J. Millis}
\affiliation{Department of Physics, Columbia University, 538 West 120th Street, New York, New York 10027, USA}
\author{A. D. Kent}
\affiliation{Department of Physics, New York University, 4 Washington Place, New York, NY 10003, USA}
\author{M. P. Sarachik}
\affiliation{Department of Physics, City College of New York, CUNY, New York, New York 10031, USA}
\author{Y. Yeshurun}
\affiliation{Department of Physics, Bar-Ilan University, Ramat-Gan 52900, Israel}
\date{\today }

\begin{abstract}
A theoretical model for  the $Mn_{12}$ acetates is derived including a 'random field' effect arising from isomer effects in some families of host acetate materials.  Estimates for the important energy scales are given. Phase boundaries are determined using a mean field approximation. Predictions for experiment are presented \end{abstract}


\maketitle

\section{Introduction}

Quantum criticality, the complex of phenomena associated with  phase changes driven at temperature $T=0$  by a variation of Hamiltonian parameters or external fields,  is a unifying theme in modern condensed matter physics. Two of the key theoretical paradigms for quantum criticality are the ``transverse field" and ``random field"  Ising models. These models are based on a set of (two-state)  Ising spins subject to an interaction $\sum_{ij}J_{ij}S_i^zS_j^z$ favoring long ranged magnetic order as well as to a transverse field $\Delta\sum_i S_i^x$ which acts to  disrupt the long ranged order by mixing the eigenstates of $S_z$ and to a random field $\sum_ih_{ran,i}S^z_i$ which acts to disrupt the long ranged order by randomly favoring different orientations of the Ising spin on different sites. 

The transverse field and random field Ising models have been the subject of extensive theoretical study, but few experimental realizations exist. One realization is the diluted antiferromagnet in a magnetic field introduced by Fishman and Aharony \cite{Fishman79}. Another realization is the $Li(Ho/Y)F_4$ system  \cite{Reich90,Bitko96} where the physics includes spin glass \cite{Reich90,Ancona-Torres08}  and random field \cite{Tabei06,Schechter08a}  effects. However in $LiHoF_4$  the nuclear hyperfine interactions play also  an important role \cite{Giraud03,Schechter05}.  Recently, a new system with the potential to exhibit  this physics has been uncovered: the $Mn_{12}$ acetates\cite{Barco03,Hill03,Barco05}.  The $Mn_{12}$ molecule is in effect a local magnetic moment of spin $S=10$, with a strong uniaxial anisotropy favoring the two states of maximal magnitude of spin projection along  a quantization axis defined by the structure of the $Mn_{12}$ molecule\cite{Friedman96,Thomas96}.  In the $Mn_{12}$ acetates the $Mn_{12}$ ions reside on the sites of  body centered tetragonal (BCT) lattice and the interspin distance is large enough that only the dipole interaction is important.  Unlike in the $LiHoF_4$ system, nuclear hyperfine fields are not believed to be important.  In a perfect $Mn$ acetate crystal the quantization axes of all of the $Mn_{12}$ units are parallel to the distinguished (c) axis of the BCT structure and a magnetic field applied in the place perpendicular to this axis is a transverse field which acts to mix the two states picked out by the uniaxial anisotropy. However, in some $Mn$ acetate materials residual orientational  disorder in the crystal structure \cite{Cornia02}, see also  \cite{Barco05,Takahashi04,Park04}  leads to a distribution of quantization axes, so a field applied perpendicular to the BCT c-axis  will have a component along the local spin quantization axis. Ref [\onlinecite{Wen09}] observed that this provides a random field of strength proportional to the applied field. The parameters are such that for experimentally realizable disorder, the random field may be the dominant non-thermal disordering effect. A random field with a different physical origin exists in diluted $Li(Ho/Y)F_4$.

The $Mn_{12}$-acetates therefore offer in principle the opportunity to study transverse-field and random-field driven quantum criticality. However spins on different sites  are separated widely enough that they interact only via the dipolar interaction. The   long range  of the  dipole interaction leads to a variety of interesting consequences for the macroscopic domain structure \cite{Garanin08,Biltmo09} but  renders the critical behavior essentially mean field-like \cite{Rechester71,Khmelnitskii71,Roussev03}. Further, as noted previously by many workers, the large spin magnitude suppresses quantum tunneling effects at small applied field, leading to an effectively frozen dynamics which limits the temperatures accessible in the present generation of experiments. Also, as will be explained below, the random field appears to have an unusual distribution, in which the random field actually vanishes at a non-neligible fraction of sites. Finally, an applied transverse field acts to cant the $Mn_{12}$ spins and this canting affects the Hamiltonian parameters.

In this paper we present a theoretical model of the $Mn_{12}$ acetates. The derivation of the Hamiltonian follows \cite{Chakraborty04,Schechter08} and our  results are qualitatively similar, but incorporates  features which are specific to the $Mn_{12}$ system.  We determine the phase boundaries using a mean-field approximation. We provide  estimates for the important energy scales; the dipolar coupling we obtain is in agreement with previous results of Garanin and Chudnovsky\cite{Garanin08}.  We compute the uniform susceptibility including the effects of the shape anisotropy, and determine the temperature range required to access the critical behavior and to determine the transitions temperatures accurately.  
\section{Hamiltonian and Energy Scales}

\subsection{Hamiltonian}

For our purposes the $Mn_{12}$ acetates may be viewed as a body centered tetragonal lattice with lattice constants $a=b\approx 17\AA$ and $c\approx 12\AA$.    These are the conventional ``cubic" lattice parameters, so the 8 nearest neighbors of  a site are at relative position ${\vec R}=0.5(\pm a {\hat x},\pm b {\hat y},\pm c {\hat z})$.  The $c/a$ ratio of the experimental samples is $\approx 0.7$ but we will also present results for other $c/a$ ratios.  Each site of the BCT lattice hosts a $Mn_{12}$ complex, which may be thought of as a single spin of magnitude $S=10$.  The separation between  $Mn_{12}$ molecules is large enough that the spins interact only via the dipolar fields.

A uniaxial anisotropy  favors spin states parallel or antiparallel to  a specific spatial direction defined by the structure of the $Mn_{12}$ molecule.  We take this direction to be the spin quantization (spin $z$) axis. In a perfect $Mn_{12}$-acetate crystal the spin quantization axis is aligned to the c axis of the BCT structure. In some crystals disorder leads to a misalignment of the quantization axis, which on site $i$ may deviate from the crystal $c$ axis by a small angle $\theta_i$. The misalignment is also characterized by an azimuthal angle $\phi_i$ which apparently \cite{Park04} takes one of a small set of discrete values.  It is of interest to consider the effects of a magnetic field applied perpendicular to the crystalline $c$ axis. We choose this direction to be the spin $x$.  Projecting the applied field onto the local quantization axis leads to a field of magnitude $H_T\equiv H\cos\theta_i\approx H$ parallel to the spin-x direction and to a field of magnitude $H_{ran,i}\equiv H\sin\theta_i \cos\phi_i \sim H\theta \cos\phi_i$ along the z direction of the spin quantization axis.  For azimuthal angle $\phi_i=0$ the energy splitting between the $S=\pm 10$ is $\Delta E =2Sg\mu_BH \theta_i$.  

The distribution of the random field depends on the distribution of angles $\theta_i,\phi_i$. Experiment \cite{Cornia02,Takahashi04,Barco05} and density functional calculations \cite{Park04} suggest that this distribution is discrete, with a fraction of order $1/4$ of the $Mn$ sites characterized by a $\theta_i=0$ and the remaining fraction by a $\theta_i$ which  density functional calculations indicate is about $0.5^\circ$  \cite{Park04} and experiment \cite{Takahashi04,Wen09} indicates is somewhat larger, $ \sim 1-2^\circ$.   The distribution is discussed in more detail in Appendix \ref{Hrand}.

These considerations lead to the following Hamiltonian
\begin{eqnarray}
H&=&H_{single-ion}+H_{intersite}
\label{H} \\
H_{single-ion}&=&-DS_z^2+g\mu_BH_TS_x
\label{Hsingleion}\\
&&\hspace{0.3in}+g\mu_BH_{ran}(h_T) S_z 
\nonumber \\
H_{intersite}&=&-\frac{(g\mu_B)^2}{2}\sum_R {\vec S}_R\cdot {\vec I}_R
\label{Hintersite}
\end{eqnarray}
with
\begin{equation}
{\vec I}_R=\sum_{R'\neq 0}
\frac{3{\vec R^{'}}\left({\vec S}_{R+R'}\cdot\vec{R}'\right)- {\vec S}_{R+R'} |\vec{R}'|^2}{ |\vec{R}'|^5}
\label{Idef}
\end{equation}
Here $R$ and $R^{'}$ label sites of a body centered tetragonal lattice and  the components of the spin operator ${\vec S}$ are matrices corresponding to the  spin $S=10$ representation of $SU(2)$. 

There are additional small terms in Eq \ref{Hsingleion} relating to quartic and higher nonlinearities, which are not important for our considerations.  We account for their effects by renormalizing the coefficient $D$ of the uniaxial anisotropy  from the measured value to $D\approx 0.548K$  to $D\approx 0.665K$ \cite{Friedman96,Thomas96,Wen09}. The uniaxial anisotropy term favors the states with $|S_z|=10$. The gap to the next lowest lying states is $19D\approx 13K$ at $H_T=0$; as we shall see, in an  applied transverse field the gap may become as small as $6T$, but this is always large compared to the energy scales relevant to phase transitions, so a reduction to an effective two-level system is generally appropriate.   

We take the free electron $g$-factor $g=2$ so that an applied transverse field of $6T$ corresponds to an energy of about $10K$, enough to rotate the spin significantly away from the $z$ axis preferred by the $D$ term.  For a tilt angle of $1^\circ$ the scale of the  random-field induced splitting $\Delta E\approx 0.25K \times H[T]$.

The strength of the dipolar interaction at the interatomic separation is set by 
\begin{equation}
E_{dip}\equiv \frac{(g\mu_B)^2S^2}{a^2c}\sim  0.078K
\label{Edip}
\end{equation}
and the geometrical factors lead to an effective spin-spin interaction scale about $5-10 \times E_{dip}$. Because this interaction scale  is much less than the splitting induced by the uniaxial anisotropy, we may focus only on the lowest-lying states of Eq \ref{Hsingleion}. We observe that because the dipole interaction varies as the cube of the distance, modestly smaller intersite distances could  dramatically increase the energy scale associated with the dipolar coupling.

\subsection{Solution of Single-Ion Hamiltonian}

\begin{figure}[t]
\includegraphics[width=0.9\columnwidth]{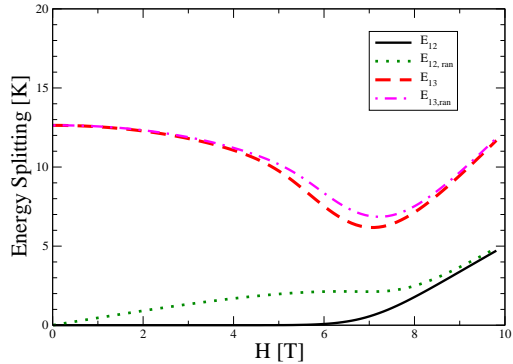}
\caption{Energy difference between ground state and first  (solid trace, black on-line and dotted trace, green on-line ) and  second (dashed line, red on-line and dash-dotted trace, magenta on-line) excited levels of $Mn_{12}$ molecule as function of applied transverse field (in Tesla), computed from solution of  Eq \ref{Hsingleion}  with (solid and dashed) and without (dotted and dashed-dotted) random field. }
\label{e13}
\end{figure}

In this subsection we present the solution of $H_{single-ion}$ as a function of  applied transverse magnetic field. The solution is straightforwardly obtained by diagonalizing Eq \ref{Hsingleion}, which is just a $21\times 21$ matrix. In the absence of applied  fields the eigenstates come in degenerate pairs. The pair with $S_z=\pm10$ are favored and the gap to the next lowest-lying pair is about $12K$.   Because the two states favored by the unixial  anisotropy are $S\pm10$ the transverse field, which couples in essence to $S_\pm$, has a highly nonlinear effect. In a simple perturbative picture $20$ applications of the operator coupling to the transverse field are required to shift the system from one ground state of $-DS_z^2$ to the other; thus at small applied fields  the splitting of the ground state doublet  induced by a small transverse field $h_T$ is $\sim (h_T/D)^{20}$ and is negligible. Above a critical value the splitting becomes important. However,  even if the splitting of the two lowest eigenvalues is negligible,  an applied transverse field  $H$ will cant the spins away from the $z$ axis, changing  the ground state wave function and the gap to the next lowest-lying states appreciably. Fig.  \ref{e13} shows the dependence on applied transverse field of the energy difference $E_{12}$ between the ground state and the lowest-lying excited state (solid trace, black on-line and dotted trace, green on-line ) and  second (dashed line, red on-line and dash-dotted trace, magenta on-line)  and the difference $E_{13}$ between the ground state and the next-lowest-lying  excited state. These differences are  computed for the model with (solid and dashed) and without (dotted and dash-dotted) random fields. We see that there is always a significant gap to the third excited level.  The minimum in the $1-3$ energy difference is a consequence of spin canting while the apparent threshold behavior of the $1-2$ energy difference is a consequence of the highly nonlinear action of $S_x$ on the ground-state doublet. 

\begin{figure}[t]
\includegraphics[width=0.85\columnwidth]{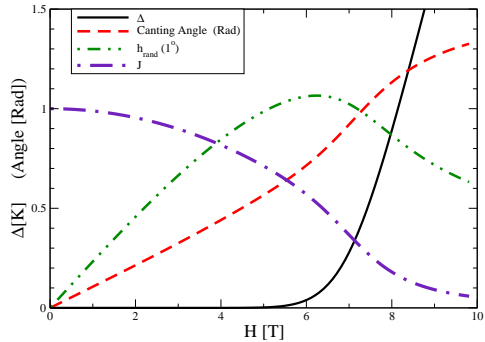}
\caption{Hamiltonian parameters for $Mn_{12}$ as function of applied transverse field (in Tesla). Solid line (black on-line) splitting $\Delta$ of two lowest levels induced by transverse field. Dash double-dotted line (green on-line): energy splitting parameter $h_{ran}$  induced by combination of applied transverse field and one degree misalignment of $Mn_{12}$ quantization axis. Dashed line (red on-line) quantum canting  angle $ArcSin[<S_x>/S]$.  Dash-dotted line (violet on-line) indicates variation with applied field of effective ferromagnetic exchange constant $J$. }
\label{splitandang}
\end{figure}

The large gap to the third excited level justifies a reduction of the Hamiltonian to a two-state system in which each site  characterized by a mean canting angle $\theta$ such that   $<S_x>(H)=S\sin\theta$.   We have computed $\theta$ from $H_{single-ion}$ for the $Mn_{12}$ parameters. The result is shown as the dashed  line (red on-line)  in Fig \ref{splitandang}. The two allowed values of $<S_z(H)>$ are $\pm \sqrt{S^2-S_x^2}=\pm S \cos\theta$. To represent this situation we introduce Ising variables $s_i$ on each site $i$ such that the spin magnitude on site $i$ is
\begin{equation}
{\vec S}(R_i)=S\left(s_i\cos\theta {\hat z}+\sin\theta {\hat x}\right)
\label{sdef}
\end{equation}

The two lowest-lying states are coupled by quantum tunneling which we parametrize by an energy  $\Delta(H)$. Further as noted above, a misalignment of the $Mn_{12}$  by a small angle $\theta_i$ leads to an effective field along the $z$ spin direction and hence a additional contribution to the  level splitting which we parametrize by an energy $h_{ran}$ which can be relatively large because of the large value of the z moment. Thus the projection of  the on-site Hamiltonian to the two lowest-lying states may be written

\begin{equation}
H(s_i)=\Delta(H)\sigma^x +h_{ran}(H)\sigma_z
\label{Hisingonsite}
\end{equation}

We define the tunnel splitting parameter $\Delta(H)$ from the energy difference $E_{12}$ of the two lowest-lying levels of $H_{single-ion}$ with no random field via $\Delta=0.5E_{12}$ and we define the random field parameter $h_{ran}$ from the difference $\Delta E_{12}$ between the splitting of the two lowest levels with and without random field, via $2\Delta E_{12}=\sqrt{\Delta^2+h_{ran}^2}-\Delta$. These energy parameters are also shown in Fig \ref{splitandang}.

\subsection{Intersite Energy}

In terms of the Ising variables $s_i$ and angle $\theta$ introduced in the previous section  the intersite energy, Eq \ref{Hintersite} becomes
\begin{equation}
H_{intersite}=-\frac{(g\mu_BS)^2}{2}\sum_R\left( s_R \cos\theta  {\bar I}^z(R)+\sin\theta {\bar I}^x(R)\right)
\label{Hintersite2}
\end{equation}
The applied magnetic field enters via the spin canting angle $\theta$  and the other terms are
\begin{eqnarray}
{\bar I}^z&=&\sum_{R^{'}\neq 0}\frac{s_{R+R^{'}}(3\cos\theta (Z^{'})^2-|R^{'}|^2)+3\sin\theta Z^{'}X^{'}}{|R^{'}|^5}
\\
{\bar I}^x&=&\sum_{R^{'}\neq 0}\frac{\sin\theta ( 3(X^{'})^2-|R^{'}|^2)+3s_{R+R^{'}}\cos\theta Z^{'}X^{'}}{|R^{'}|^5}
\end{eqnarray}

Thus the intersite energy is the sum of three terms: a true interaction which is bilinear in the Ising variables, a term linear in the Ising variables which expresses the $z$ component of the magnetic field arising (via the dipolar interaction) from the $x$ direction spin polarization, and a term which is independent of the Ising variables, which we neglect henceforth. Recalling the dipole energy $E_{dip}$ (Eq \ref{Edip}) and  the unit cell volume $V_{cell}=a^2c/2$  we have (similar results were found for $LiHoF_4$ by Chakraborty et. al. \cite{Chakraborty04} and for $Mn_{12}$ acetate by Garanin and Chudnovsky \cite{Garanin08}) 
\begin{eqnarray}
H_{intersite}=H_{interaction}+H_{orientation}
\end{eqnarray}
\begin{equation}
H_{interaction}=-\frac{E_{dip}\cos^2\theta}{2} \sum_{R,R^{'}\neq R}s_R s_{R^{'}} K({\bf R}-{\bf R}^{'})
\label{Hintersite2}
\end{equation}
\begin{equation}
H_{orientation}=-E_{dip}\cos\theta \sin\theta \left(2V_{cell}\right)  \sum_Rs_R H_{or}({\bf R})
\label{Horientation}
\end{equation}
with 
\begin{equation}
K({\bf R})=2V_{cell} \frac{2Z^2-X^2-Y^2}{|{\bf R}|^5}
\end{equation}
and
\begin{equation}
H_{or}({\bf R})=\sum_{R^{'}\neq R} \frac{3(Z-Z^{'})(X-X^{'})}{|R-R^{'}|^5}
\end{equation}
$H_{or}cos\theta$ is shown in Appendix \ref{Ewaldmeanfield} to be just the $z$ component of the demagnetization field induced by the $x-$direction spin polarization caused by spin canting. This vanishes for a properly oriented ellipsoidal sample and is generally expected to be small. It will be neglected henceforth. 

As will be seen below, an appropriate measure of the interaction strength is the ferromagnetic exchange constant which for the $Mn_{12}$ acetate $c/a$ ratio is   $J=(2.46+ 8\pi /3)E_{dipole}$ (this number is in good agreement with the value previously established by Garanin and Chudnovsky \cite{Garanin08}). $Jcos^2\theta$ is also plotted in Fig \ref{splitandang}. In the remainder of this paper we analyse the physics implied by $H_{Ising}$ (Eq \ref{Hisingonsite}), $H_{interaction}$ (Eq \ref{Hintersite2}) and $H_{orientation}$ (Eq \ref{Horientation}).  

\section{Mean Field Theory}
\subsection{Formalism}
In mean field theory one assumes that  each site is an independent spin problem specified by the Hamiltonian
\begin{equation}
H_{MF}(R)=\Delta \sigma_x -h(R)_{tot}\sigma_z
\label{HMF}
\end{equation}
with $z$ direction magnetic  field  the sum of the random field term, any externally applied field, and a contribution coming from the polarizations of the other spins:
\begin{equation}
h(R)_{tot}=h_{ran}(R)+h_{app}(R)+h_{eff}(\{<s_{R^{'}\neq R}>)
\end{equation}

Eq \ref{HMF} implies that the expectation value of the Ising spin operator on site $R$ is
\begin{equation}
<s_R>=\frac{h_{tot}(R)}{\sqrt{h_{tot}^2(R)+\Delta^2}}tanh\frac{\sqrt{h^2_{tot}(R)+\Delta^2}}{T}
\label{savg}
\end{equation}

For ordering at a wavevector ${\vec Q}$ we have $<s_{{\bf R}}>=Re <s>e^{i{\bf Q}\cdot {\bf R}}$ and  
\begin{equation}
h_{eff}=J_{eff}(Q)<s>
\label{mfsce}
\end{equation}
with
\begin{equation}
J_{eff}(Q)=E_{dip}\cos^2\theta  \left(2V_{cell}\right) \sum^{{'}}_{R\neq 0} e^{i{\vec  Q}\cdot {\vec R}}  \frac{3Z^2-R^2}{|R|^5}
\label{heff}
\end{equation}
The prime on the sum denotes the restriction that the site $R$ must be within the sample volume.  

Within mean field theory  the effective field $h_{eff}$ determined by putting the expectation values computed from Eq \ref{HMF} back into the equation for  $h_{eff}$ and requiring self-consistency. A magnetic state is found when   self consistency occurs for $<s>\neq 0$ in vanishing applied $z$-direction field. A second order magnetic phase boundary  is defined by the temperature at which Eqs \ref{savg},\ref{mfsce} are satisfied by an infinitesimal $<s>$ at vanishing applied $z$-direction field. The nature of the phase is determined by $Q$ which maximizes $J(Q)$.

Appendix  \ref{Ewaldmeanfield} presents the evaluation of $J(Q)$. For the ferromagnetic case $Q=0$ careful attention must be paid to the long range of the dipole interaction because the sum is only conditionally convergent; for non-zero $Q$ the  complication does not arise.  For the ferromagnetic case we find
\begin{equation}
J_F=E_{dip}\cos^2\theta \left(2J_{SR}\left(\frac{c}{a}\right)+\frac{8\pi}{3}-2\Lambda\right)
\label{Jfinal1}
\end{equation}
Here $J_{SR}$ comes from short ranged physics and depends on the details of the crystal structure including (for the BCT lattice) the $c/a$ ratio. For the $c/a=0.7$ relevant to the $Mn_{12}$ acetates we find $J_{SR}(0.7)\approx 1.23$. The second term comes from the long ranged part of the dipole interaction and is independent of the specifics of the crystal structure or the over-all shape of the sample. This term is in effect a long-ranged interaction, which justifies the use of a mean field theory and is of course absent in the antiferromagnetic case.   $\Lambda$ is  the demagnetization factor, which is non-negative but tends to zero  for a prolate crystal highly elongated in the direction parallel to the applied field. Eq \ref{Jfinal1} was derived on the assumption of a uniform ferromagnetic state. In a crystal which is not highly prolate, the ordered state will have a domain structure consisting of domains highly elongated along $z$ to minimize the demagnetization factor; thus the ferromagnetic  transition temperature is determined by Eq \ref{Jfinal1} with $\Lambda=0$.  

\begin{figure}[t]
\includegraphics[width=0.9\columnwidth]{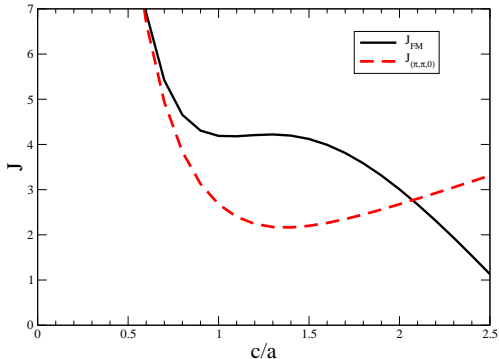}
\caption{Exchange constant appearing in mean field theory of ferromagnet (solid line, black on-line, obtained from Eq \ref{Jfinal1} with demagnetization factor $\Lambda=0$) and $(\pi,\pi,0)$ antiferromagnet (dashed line, red on-line) computed as function of $c/a$ ratio of body centered tetragonal lattice and expressed in units of $2E_{dip}\cos^2\theta$. }
\label{ferrovsaf}
\end{figure}

We have also studied the $Q$ dependence. We find that the largest exchange constants are for $Q=0$ and for    ${\bf Q}=(\pi,\pm\pi,0)$. When translated into a real-space picture of sites on the BCT lattice the  ${\bf Q}=(\pi,\pm\pi,0)$  state corresponds to  ferromagnetic sheets oriented perpendicular to the basal plane of the BCT and extending along $(1,\pm1)$ directions of the simple cubic lattice from which the BCT lattice is constructed. 

\begin{figure}[tbph]
\includegraphics[width=0.9\columnwidth]{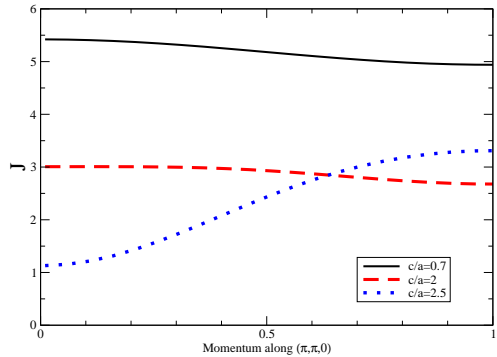}
\caption{Dependence of mean field exchange constant (measured in units of $E_{dip}\cos^2\theta)$ on momentum in $xy$ plane for body centered tetragonal lattice at three representative values of the $c/a$ ratio. }
\label{qdependence}
\end{figure}

Energy differences depend sensitively on the $c/a$ ratio and can lead to a change of ordering pattern.  For $c/a\lesssim 2.5$ the ferromagnetic state is favored; for $c/a\gtrsim 2.5$ the $(\pi,\pi,0)$ antiferromagnet has the lowest energy. The dependence on the $c/a$ ratio  of the exchange constant characterizing these two states is shown in Fig \ref{ferrovsaf}. We find that for $c/a\lesssim 0.5$ the ferromagnet and the $(\pi,\pi,0)$ antiferromagnet become extremely close in energy while for $c/a \gtrsim 2.2$ the state preferred within mean field theory is the $(\pi,\pi,0)$ antiferromagnet, with the $(\pi,\pi,\pi)$ antiferromagnet becoming extremely close in energy as $c/a$ is further increased. The proximity of these other states may be important in the random field case, as it is possible to imagine that particular configurations of the random field might locally favor one or the other of the states. The dependence of the mean field exchange constant  on ordering wavevector along the basal plan Brillouin zone diagonal is shown in Fig \ref{qdependence}.

\subsection{Phase boundaries: ferromagnetic case, $c/a=0.7$}

The ferromagnetic phase boundary is determined in mean field theory by linearizing Eq \ref{savg} in $s$ and then seeking the temperature at which Eq \ref{mfsce} is satisfied. As noted above, in general the transition is to a state with domains highly elongated along $z$ so that in determining the transition temperature we evaluate $J$ with $\Lambda=0$. Denoting by angle brackets the average over sites in the system the equation for the Curie temperature $T_c$ is
\begin{eqnarray}
1&=&\left<\frac{J\Delta^2}{(h_{ran,i}^2+\Delta^2)^{3/2}}\tanh\frac{\sqrt{h_{ran,i}^2
+\Delta^2}}{T_c}\right>
\nonumber \\
&+&\left<\frac{h_{ran,i}^2 J}{T_c(h_{ran,i}^2+\Delta^2)}\cosh^{-2}\frac{\sqrt{h_{ran,i}^2+\Delta^2}}{T_c}
\right>
\label{Tc}
\end{eqnarray}
If the randomness vanishes the mean field equation becomes

\begin{equation}
1=\frac{J}{\Delta}tanh\frac{\Delta}{T_c}
\label{MFpure}
\end{equation}

The quantum critical point at which the mean field transition vanishes is $J=\Delta$; the corresponding field can be read off from Fig \ref{splitandang}.

The random field case is more involved. Fig \ref{splitandang} indicates that  for the experimentally measured tilt angles the random field scale becomes comarable to the basic exchange scale while the tunnel splitting $\Delta$ is still very small. Neglecting $\Delta$ in Eq \ref{Tc} we obtain

\begin{equation}
1=\left<\frac{J}{T_ccosh^2\frac{|h_{ran,i}|}{T_c}}\right>
\label{MFrandom}
\end{equation}
The physics of this equation is straightforward: as $T$ is decreased below $h_{ran,i}$ the $cosh^{-2}$ term becomes negligible so these sites drop out of the mean field equation.   As the random field is increased a higher and higher fraction of sites have dropped out of the mean field equation at any given temperature  so the transition temperature drops. If the distribution of random fields were continuous, Eq \ref{MFrandom} would lead to a quantum critical point at which $T_c$ vanished even without a tunnel splitting. However, in the actual materials the distribution of random fields is apparently \cite{Cornia02,Takahashi04,Park04} such that a non-negligible fraction of sites experience zero random field (see Appendix \ref{Hrand}); this fraction can (within mean field theory) sustain an ordered state which is only suppressed by tunnel splitting.  

\begin{figure}[t]
\includegraphics[width=0.9\columnwidth]{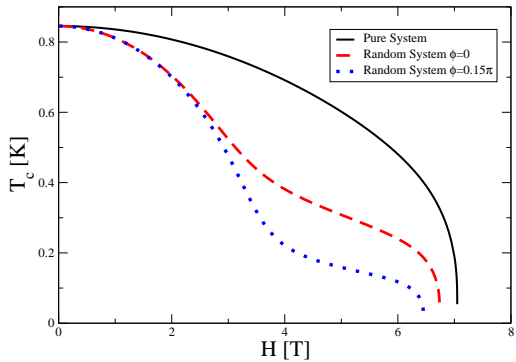}
\caption{Transition temperatures for pure (solid line, black on-line) and two random-field cases: applied field aligned along crystal $x$ axis  (dashed line, red on-line) and aligned at an angle of $0.15\pi$ with respect to the crystal $x$ axis. $Mn_{12}$ parameters as discussed above are used, with polar tilt angle $\theta_i$ taken to be $1^\circ$ }
\label{phaseboundaries}
\end{figure}

We have used the field dependence of the exchange, tunnel splitting and random field energies to compute the phase boundaries implied by Eq \ref{Tc}.  In the random case we have used the distribution function shown in Appendix \ref{Hrand}.  The results are shown in Fig \ref{phaseboundaries}. In the pure system we observe a roughly mean-field-like curve, with however a very steep rise near the endpoint which is related to the rapid onset of the tunnelling term in the Hamiltonian. In the random field case we observe an initial drop in transition temperature related to freezing out of spins quenched by the random field. The field scale for the initial drop is set by the polar tilt angle $\theta_i$ which in Fig \ref{phaseboundaries} has been set to $\theta_i=1^\circ$. If the typical $\theta_i$ for sites with tilt were a factor of two smaller, the field scale at which the drop occurs would be about a factor of two larger (with small corrections relating to the effect of the over-all canting on the basic exchange energies). For  larger fields  ($H>2T$ for the parameters used in Fig \ref{phaseboundaries}) the calculated behavior  controlled by the sites where the random field vanishes; the number of these sites depends on the orientation of the field relative to the molecule axis defined by the coordinate $\phi$. If the applied field is aligned along the direction $\phi=0$ then some fraction of tilted sites are characterized by $\phi=\pm \pi/2$ and thus vanishing random field; for a generic angle of applied field, all of the tilted sites are subject to some random field. Remarkably, a phase diagram with a shape very similar to that found in our random field case was computed for diluted $LiHoF_4$ \cite{Schechter08} although in this  case the structure is due to the interplay of hyperfine and dipolar interactions.

\section{Spin-spin interaction}

It is interesting to consider the spin-spin interaction term in a formally infinite system, so that momentum is a good quantum number and  $H_{interaction}$ becomes (with the use of Ewald summation techniques along the lines of Appendix \ref{Ewaldmeanfield}  to Fourier transform the dipole interaction) 
\begin{equation}
H_{interaction}=\frac{1}{2}\frac{8\pi E_{dip}\cos^2\theta}{3}V_{cell}\int \frac{d^3k}{(2\pi)^3}\zeta(k)s_ks_{-k}
\label{Hkspace}
\end{equation}

with 

\begin{equation}
\zeta(k)=\frac{3k_z^2}{k^2}+\zeta_{SR}(k)
\label{phikdef}
\end{equation}
with
\begin{eqnarray}
\zeta_{SR}(k)&=&\left(e^{-\frac{k^2}{4q^2}}-1\right)\frac{k^2-3k_z^2}{k^2}
 \\
&&+\sum_{G\neq 0}  e^{-\frac{({\vec k}+{\vec G})^2}{4q^2}} \frac{(k+G)^2-3(k+G)_z^2}{(k+G)^2}
\nonumber \\
&&+\frac{1}{\pi^{3/2}}\left(2V_{cell}\right)\sum_{R\neq 0}I(R)e^{i{\vec k}\cdot{\vec R}} \left(3Z^2-R^2\right)
\nonumber
\label{JSR}
\end{eqnarray}
with
\begin{equation}
I(R)=\int_q^\infty d\kappa \kappa^4 e^{-\kappa^2 R^2}
\end{equation}

Here $G$ is a reciprocal lattice vector and  $q$ is the separation parameter used to effect the Ewald summation; the results are independent of the value of $q$ chosen, but values of $q$ between $1.5$ and $2.5$ seem to lead to the most rapid convergence. This result applies only for a ferromagnetic ground state and for  $k$ large compared to the inverse of the domain size. 

\begin{figure}[t]
\includegraphics[width=0.9\columnwidth]{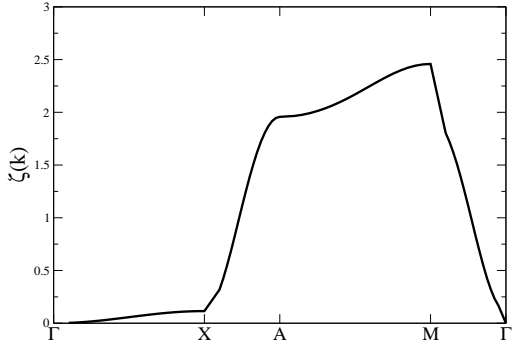}
\caption{Variation with wavevector of magnetic energy for anisotropy ratio $c/a=0.7$. Here $\Gamma=(0,0,0)$, $X=(\pi,\pi,0)$, $A=(\pi,\pi,\pi)$ and $M=(0,0,\pi)$. }
\label{exchange}
\end{figure}

The first term in Eq \ref{phikdef} is independent of the magnitude of $k$ but strongly dependent on the ratio $k_z/k$. It is independent of any of the atomic-scale details of the material but  carries information about the shape anisotropy of the sample and it favors states with $k_z=0$ but for $k_z=0$ is independent of the wave vector in the $xy$ plane.  The remaining terms are not singular as $k\rightarrow 0$ and depend on the magnitude of $k$ and on the crystal structure.  The variation of the energy as momentum is varied through the Brillouin zone of the BCT lattice for $c/a=0.7$  is plotted in Fig \ref{exchange}. As the lattice anisotropy $c/a$ is decreased below $1$ the variation in energy across the $k_z=0$ plane rapidly decreases.

\section{Susceptibility}

An important experimental probe is the magnetic susceptibility.  At present \cite{Wen09} this can be carried out for $Mn_{12}$ only at relatively high temperatures (at least for small applied transverse fields) because the small value of the tunnel amplitude $\Delta$ means that the system drops out of equilibrium as the temperature is decreased.  Phase diagrams have been inferred from a Curie-Weiss extrapolation of the measured $\chi$. 

We calculate the susceptibility by writing the mean field equations in the presence of a small probe field. Expanding  for small probe field and  small magnetization we find (the canting angle enters the equations because it determines the magnitude of the Ising spin):
\begin{equation}
\chi^{-1}(H,T)=\frac{1-Jcos^2\theta( I_1(H,T)+ I_2(H,T))}{\left(I_1(H,T)+I_2(H,T)\right)\cos^2(\theta)}
\label{chi}
\end{equation}
with the $\cos\theta$ factor expressing the canting of the spins in the transverse applied field,
\begin{eqnarray}
I_1&=&\left<\frac{\Delta^2}{\left(h_{ran}^2+\Delta^2\right)^{3/2}} tanh\left[\frac{\sqrt{h_{ran}^2+\Delta^2}}{T}\right]\right>
\\
I_2&=&\left<\frac{h_{ran}^2}{T\left(h_{ran}^2+\Delta^2 \right)} sech^2\left[\frac{\sqrt{h_{ran}^2+\Delta^2}}{T}\right] \right>
\end{eqnarray}
and the angle brackets again representing an average over the random field.  The susceptibility is a measurement of the uniform magnetization induced by a uniform field, and as such is substantially affected by shape anisotropy effects which as noted above affect the domain structure of the ordered state but not the ordering temperature. One should therefore include the demagnetization contribution $\Lambda$ in the exchange constant $J$.

\begin{figure}[t]
\includegraphics[width=0.9\columnwidth]{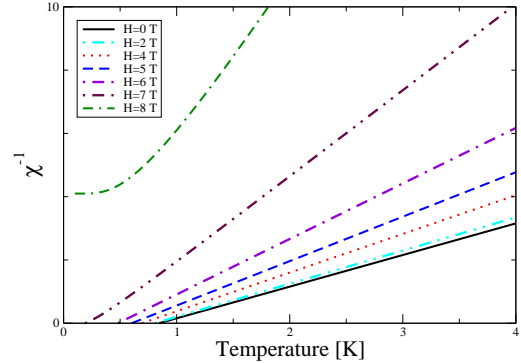}
\includegraphics[width=0.9\columnwidth]{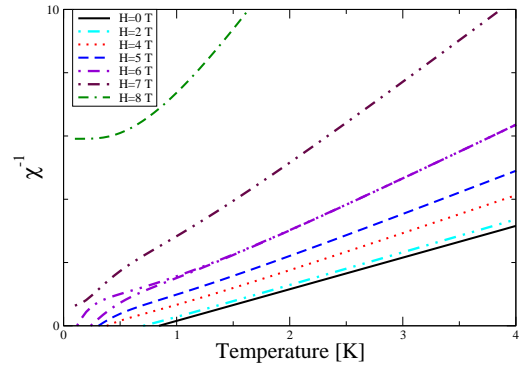}
\caption{Inverse susceptibility calculated for highly prolate (demagnetization factor $\Lambda=0$)  $Mn_{12}$ sample using parameters as described in the text. Upper panel: pure system. Lower panel: random system. Random system: isomer azimuthal orientation angle $\phi=0$ except for $H=6T$ where both $\phi=0$ (lower curve) and $\phi=0.15\pi$ (upper curve) are shown. }
\label{chiinv}
\end{figure}

Some insight comes from considering the susceptibility in the high temperature limit. Expanding Eq \ref{chi} yields
\begin{equation}
\chi^{-1}(H,T)=\frac{T}{\cos^2\theta}-J+\frac{\frac{1}{3}\Delta^2+<h_{ran}^2>}{T\cos^2\theta}+{\cal O}T^{-2}
\label{chihigh}
\end{equation}
Thus the first correction to the Curie-Weiss behavior is an upward curvature whose amplitude depends on the tunnel splitting and on the average strength of the random field.  However, when temperature becomes low enough that for a given site $h_{ran,i}$ becomes greater than $T$, then the contribution proportional to $I_2$ vanishes and the contribution proportional to $I_1$ becomes equal to the sign of $h_{ran}$ and vanishes on averaging; thus mathematically this site drops out of the mean field equation for the susceptibility.

Fig \ref{chiinv}  shows  results for a highly prolate sample (needle-like, elongated along $z$) for which $\Lambda=0$. In the pure case (top panel) the field dependence for small fields is entirely due to the field dependence of the basic exchange constant; by contrast in the random case (lower panel) there is additional field dependence is due to the random field.  The random field case has structure at low temperatures, caused by the physics discussed above: a non-negligible fraction of the sites have a vanishing or very small random field; the sites are able (at small fields and low temperatures) to order. Note that the dependence on alignment of the applied field and the sample $x$ axis is only important very near to the ordering transition, as can be seen by comparing the two curves shown for $H=6T$.

\begin{figure}[t]
\includegraphics[width=0.9\columnwidth]{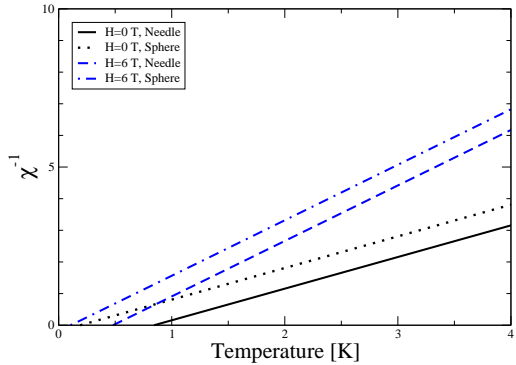}
\caption{Temperature dependence of inverse susceptibility of non-disordered samples at two applied transverse fields for highly prolate (needle-shaped) and spherical samples. }
\label{chishape}
\end{figure}

Fig \ref{chishape} compares the calculated susceptibility for needle like ($\Lambda=0$; no demagnetization factor) and spherical ($\Lambda=4\pi/3$; demagnetization factor cancels `Clausius-Mosotti' $4\pi/3$ factor) samples with no random fields at two representative applied fields. The $\chi^{-1}=0$ intercept of the curves reveals the `short ranged' part of the interaction. The effect of the shape anisotropy is evident.

\section{Conclusion}
A model of spins with a uniaxial anisotropy interacting via the dipolar interaction has been presented and applied to data on $Mn_{12}$-acetates. The model includes  physics specifically relevant to $Mn_{12}$, in particular the  crystal structure and a random field arising from an isomer structure of the some of the host acetate materials.  The model is similar to that previously formulated for $LiHoF_4$ \cite{Bitko96,Chakraborty04,Schechter08} although the leading term in the random field has a different physical origin in $Mn_{12}$ than in $LiHoF_4$. One should note  that although $LiHoF_4$ crystallizes in a body centered tetragonal structure, the $LiHoF_4$ unit cell has a more complicated structure than that of  $Mn_{12}$ (including 4 atoms per unit cell to the 2 in $Mn_{12}$) so the dependence of the exchange constants on the $c/a$ ratio, for example, will be different in the two systems.   Also, $Mn_{12}$  is simpler than $LiHoF_4$ because the nuclear hyperfine field plays no role in $Mn_{12}$.

We presented estimates of the relevant energy scales. The phase diagram and susceptibility were computed in a mean field approximation. In the random field case the plot of transition temperature versus transverse field has an unusual structure arising from a particular feature of the  theoretically proposed distribution of random fields. The random fields arise from a random distribution of isomers of the acetate molecules which host the $Mn_{12}$ ions. The distribution is such that a fraction (either $1/4$ or $1/2$) of the sites feel no on-site random field, while the remainder are subject to a random field, of strength proportional to the transverse applied field.   The large fraction of sites on which the random field vanishes leads to a `foot', extending to higher fields,  in the calculated transition temperature versus applied transverse field curve.  The strength of  the random field, and therefore the field and temperature  regime in which the unusual structure of the phase boundary  becomes apparent, depends on the mean tilt angle. Theory and experiment differ on what is the best value of this angle to use; however for all reasonable parameters it appears  that  measurements at temperatures below $1K$ will be required to unravel the nature of this transition.

An important experimental probe  of the $Mn_{12}$ system is the magnetic susceptibility. These measurements are at present limited to relatively high temperatures because (especially at applied fields less than a few $T$) the  small value of the effective tunnelling amplitude means that the system has difficulty equilibrating at low $T$. It is therefore of interest to consider the information which may be obtained from measurements of the susceptibility  at higher temperatures, above the actual ordering temperature. Caution in choosing a temperature range  is suggested because our results also indicate that  particularly for fields $\sim 5-7T$ higher excited states (not included in the present calculations) may start to play a role at temperatures $\sim 5K$ or greater. We find that  at temperatures well above any ordering temperatures the leading effect of the random field is an upward deviation of the inverse susceptibility,  $\sim 1/T$, from the Curie Weiss behavior. The coefficient of the deviation gives the mean square amplitude of the random field.  

Our phase diagrams and susceptibilities are obtained from mean field theory, which should provide a reasonable estimate of the scales but which is not highly accurate even with the long range of the dipole interaction \cite{Chakraborty04,Biltmo09}.  In particular, the mean field theory captures the leading effect of the random field, namely that the spins on sites where the random field is non-zero   are 'slaved' to the random field and thus drop out of the mean field equation, leading to an effective dilution that reduces the transition temperature.  However, the mean field approximation neglects the dipolar coupling between  the frozen moments on the misaligned sites and the potentially  ferromagnetically ordered spins on the aligned sites. This coupling leads to an additional random field which averages to zero but has a typical value of the order of the basic exchange constant \cite{Fishman79,Tabei06,Schechter08a}. It also neglects the possibility of other spin glass state or frozen moment states at low dilutions (such states are believed to occur for dilutions greater  than about $80\%$ in $LiHoF_4$ \cite{Reich90,Ancona-Torres08}). A rough estimate indicates that in the $50\%$ dilution case (occurring if the transverse field is aligned along an azimuthal symmetry axis) the root mean square value of the dipolar-induced random field is less than the mean field associated with ferromagnetic ordering, but for a general azimuthal field alignment the root mean square value of the dipolar-induced random field may be larger than the ferromagnetic mean field.   

Important topics for future investigation are  a better characterization of the random fields,  a more theoretically rigorous investigation of the model we have defined  (along the lines of \cite{Roussev03,Biltmo09}), including an investigation of the interplay between  the random field and the mean field-like interaction, and an investigation of materials in which quantum fluctuations are larger, enabling measurements down to lower temperatures. Forming crystals with a smaller lattice constant and different $c/a$ ratios would  also be of considerable interest because they would vary the dipolar interaction. 

{\it Acknowledgements} We thank D. Garanin, M. Schechter and P. Stamp for helpful conversations. A.J.M. thanks the NYU physics department for hospitality and was supported by NSF-DMR--0705847. A.D.K. acknowledges support by NSF-DMR-0506946 and ARO W911NF-08-1-0364, M.P.S. acknowledges support by NSF-DMR-0451605 and Y.Y. acknowledges support of the Deutsche Forschungsgemeinschaft through a DIP project and thanks the NYU and CUNY  physics departments for hospitality.

\begin{appendix}

\section{Distribution of Random Fields \label{Hrand}}

The distribution of random fields arises physically from a distribution of isomers of the host acetate material \cite{Park04}. This distribution is believed \cite{Park04} to be such that $1/4$ of the molecules are untilted; the remaining $3/4$ are tilted by angles $\theta_i$ which density functional calculations indicate are $0.4^\circ$ or $0.5^\circ$ (depending on the isomer) and which experiment indicates is somewhat larger. In our modelling we neglect the difference between $0.4$ and $0.5$ and characterize the tilted isomers by an angle $\theta_0$ which in our numerical calculations is taken to be $1^\circ$. The effects we consider are linear in $\theta_0$ so our results may easily be rescaled to other values of $\theta_0$. The distribution of azimuthal angles is somewhat involved, and is listed in the Table. From the angles we may compute the random field  $h_r=g\mu_BH \sin \theta_0 \cos(\phi-\phi_H)$ (note we have included an angle $\phi_H$ expressing the (experimentally unknown) angle of the applied field with respect to the crystalline $x$ axis). 

\begin{center}
\begin{table}
\begin{tabular}{c|ccccccccc|}
i&1&2&3&4&5&6&7&8&9\\
$\theta_i$ & 0 & $\theta_0$  & $\theta_0$  & $\theta_0$  & $\theta_0$  & $\theta_0$  & $\theta_0$  & $\theta_0$&$\theta_0$  \\
$\phi_i$ & -- & 0 &  $\frac{\pi}{4}$  & $\frac{\pi}{2}$  & $\frac{3\pi}{4}$  & $\pi$ & $\frac{5\pi}{4}$ &$\frac{3\pi}{2}$&$\frac{7\pi}{4}$   \\
\hline
$\frac{h_i}{h_{ran}}$ & 0 & 1 & $\frac{1}{\sqrt{2}}$ & 0 & - $\frac{1}{\sqrt{2}}$ & -1 & - $\frac{1}{\sqrt{2}}$& 0&$\frac{1}{\sqrt{2}}$\\
$P_i$ & $ \frac{1}{4}$ &  $\frac{1}{8}$& $\frac{1}{16}$ &  $\frac{1}{8}$ &  $\frac{1}{16}$ &  $\frac{1}{8}$ &  $\frac{1}{16}$ &  $\frac{1}{8}$ &  $\frac{1}{16}$\\
\hline
\end{tabular}
\caption{Table of values  of polar $\theta$ and azimuthal $\phi$ angles along with random field (expressed as a fraction of $h_{ran}=g\mu_BH\sin\theta_0$ for field directed along crystal $x$ axis) and probability of occurrence for isomer $i$   host molecule in $Mn_{12}$-acetate crystals }
\end{table}
\end{center}

\section{Ewald summation and the mean field interactions  \label{Ewaldmeanfield}}

This Appendix uses Ewald summation arguments to perform the sums needed for the mean field theory of the dipolar Ising magnet.  We need to evaluate 
\begin{equation}
H_{eff}(R)=V_{cell}\sum_{R^{'}\neq R}  \frac{3(Z-Z^{'})^2-|R-R^{'}|^2}{|R-R^{'}|^5}M(R^{'})
\label{eshapeaniso}
\end{equation}

Here we find it convenient to allow the sum over $R^{'}$ to range over infinite space, and take the magnetization  $M(R^{'})=0$ for $R^{'}$ outside the sample.

We now introduce an arbitrary separation length $\xi$ which is large compared to a lattice constant but small compared to the system size and introduce kernels $K_{LR}$ and $K_{SR}|$ with 
\begin{eqnarray}
K_{LR}(R)&=&\left(1-e^{-\frac{R^2}{\xi^2}}\right)\left(\frac{3Z^2-R^2}{R^5}\right)
\label{K1def}\\
K_{SR}(R)&=&e^{-\frac{R^2}{\xi^2}}\left(\frac{3Z^2-R^2}{R^5}\right)
\label{K2def}
\end{eqnarray}

The term in Eq \ref{eshapeaniso} involving $K_{SR}$ is (up to boundary terms) local and independent of the shape of the system, however it does depend on the crystal structure including,  for a BCT lattice, the  $c/a$ ratio. Assuming that $M$ varies slowly on the scale of $\xi$ it  gives  a contribution
\begin{equation}
H_{eff}^{short}(R)=M(R)V_{cell}\sum_{R\neq 0}K_{SR}(R)
\label{E2def}
\end{equation}

We write $R^{-5}=8/(3\sqrt{\pi})\int d\kappa \kappa^4 e^{-\kappa^2 R^2}$ and split the integral into two parts so
\begin{eqnarray}
\sum_{R\neq 0}K_{SR}(R)&=&K^q_1+K^q_2
\label{split}
\end{eqnarray}
\begin{equation}
K^q_1=\frac{8}{3\pi^{1/2}}\sum_{R\neq 0}\int_0^q d\kappa \kappa^4 e^{-\kappa^2 R^2} e^{-\frac{R^2}{\xi^2}} \left(3Z^2-R^2\right)
\label{phi1def}
\end{equation}
\begin{equation}
K^q_2=\frac{8}{3\pi^{1/2}}\sum_{R\neq 0}\int_q^\infty d\kappa \kappa^4 e^{-\kappa^2 R^2} e^{-\frac{R^2}{\xi^2}} \left(3Z^2-R^2\right)
\label{phi2def}
\end{equation}

$K^q_2$ is evaluated directly.  In $K^q_1$ the summand vanishes at $R=0$ so the sum may be extended to include this term.  We introduce $1=\int d^3r \delta(r-R)$ and $\sum_R\delta(r-R)=V_{cell}^{-1}\sum_Ge^{i{\vec G}\cdot{\vec R}}$ to obtain
\begin{equation}
K^q_1=\frac{8}{3\pi^{1/2}}\sum_G\int d^3r \int_0^q \frac{d\kappa \kappa^4}{V_{cell}} e^{-(\kappa^2+\xi^{-2}) r^2+i{\vec G}\cdot{\vec r}} \left(3z^2-r^2\right)
\end{equation}
The term with $G=0$ vanishes on integration over the directions of $r$. In the $G\neq 0$ terms we shift ${\vec r}\rightarrow {\vec r}+i\frac{{\vec k}+{\vec G}}{2\kappa^2}$ and perform the $r$ intergral, obtaining
\begin{equation}
K^q_1=-\frac{8\pi}{3V_{cell}}\sum_{G\neq 0} \int_0^q \frac{\kappa^4d\kappa \left(3(G)_z^2-(G)^2\right)}{(\kappa^2+\xi^{-2})^{7/2}} e^{-\frac{{\vec G}^2}{4(\kappa^2+\xi^{-2})}} 
\end{equation}
The $\kappa$ integral is dominated by $\kappa\sim G$ so we may set $\xi=0$ and perform $\kappa$  integral obtaining finally
\begin{eqnarray}
K^q_1&=&-\frac{4\pi}{3V_{cell}}\sum_{G\neq 0}  e^{-\frac{G^2}{4q^2}} \frac{3G_z^2-G^2}{G^2}
\label{K1answer}
\\
K^q_2&=&\frac{8}{3\sqrt{\pi}}\sum_{R\neq 0}\int_q^\infty d\kappa \kappa^4 e^{-\kappa^2 R^2}  \left(3Z^2-R^2\right)
\label{K2answer}
\end{eqnarray}

We have evaluated $K_{SR}$ from  Eqs \ref{K1answer},\ref{K2answer}.

To analyse $K_1$ we introduce $1=\int d^3r^{'}\delta(r^{'}-R^{'})$ and $V_{cell}\delta(r^{'}-R^{'})=\sum_Ge^{i{\vec G }\cdot {\vec r}^{'}}$ and note that the slow variation means that only the terms with $G=0$ contribute so that 
\begin{eqnarray}
H_{eff}^1(r)&=&\int d^3r^{'} \frac{3(z-z^{'})^2-(r-r^{'})^2}{|r-r^{'}|^5}
\nonumber \\
&&\times \left(1-e^{-\frac{|r-r^{'}|^2}{\xi^2}}\right)M(r^{'})
\label{Heff1}
\end{eqnarray}
 We now observe that
\begin{equation}
\frac{3(z-z^{'})^2-(r-r^{'})^2}{|r-r^{'}|^5}=-\partial_z\partial_{z^{'}}\frac{1}{|r-r^{'}|}
\end{equation}
and integrate by parts,  so that 
\begin{equation}
H_{eff}^1(r)=\partial_z \int d^3r^{'} \frac{1}{|r-r^{'}|}\partial_{z^{'}}\left(\left(1-e^{-\frac{|r-r^{'}|^2}{\xi^2}}\right)M(r^{'})\right)
\label{integraleqn}
\end{equation}
Eq \ref{integraleqn} may be recast in terms of a differential equation for the  magnetic potential $\Phi_M$  which is related to the  demagnetizing field  $H$ by ${\vec H}=-\nabla \Phi_M$.  Let us define the general function $\Phi_M({\bar r};r)$ via
\begin{equation}
\Phi_M({\bar r};r)= -\int d^3r^{'} \frac{\partial_{z^{'}}\left(\left(1-e^{-\frac{|r-r^{'}|^2}{\xi^2}}\right)M(r^{'})\right)}{|{\bar r}+r-r^{'}|}
\label{integraleqn2}
\end{equation}
so that $H_{eff}(r)=-\partial_{{\bar z}}\Phi_M({\bar r};r)|_{{\bar r}=0}$. We now shift the origin of the $r^{'}$ integral to $r$ and write the differential form of Eq \ref{integraleqn2} as
\begin{equation}
\nabla^2\Phi_M({\bar r};r)=4\pi \partial_{{\bar z}}\left(\left(1-e^{-\frac{|{\bar r}|^2}{\xi^2}}\right)M({\bar r})\right)
\label{diffeq}
\end{equation}
We must solve Eq \ref{diffeq} for $\Phi_M({\bar r})$, take the derivative and evaluate the result at ${\bar r}=0$. The source term has two contributions; one is from $\partial M$ which is non-vanishing only at the sample boundary (note that in our coordinate system this depends upon $r$ and that for points in the interior of the sample we may neglect the exponential). This gives us the usual demagnetization field.  In general the demagnetization field is complicated but for uniformly magnetized ellipsoidal shaped samples with the field applied along a symmetry axis we have
\begin{equation}
H_{demag}=-\Lambda M
\end{equation}
with $\Lambda$ ranging from $4\pi/3$ for a sphere to  $0$ for  'needle-like' samples very elongated in the direction parallel to the field.

For the second term we may  take $M$ to be constant. We then require the solution of 
\begin{equation}
\nabla^2\Phi_M({\bar r})=8M\pi \frac{{\bar z}}{\xi^2} e^{-\frac{|{\bar r}|^2}{\xi^2}}
\end{equation}
This is most easily solved in Fourier space as
\begin{equation}
\Phi_M(k)=-4M\pi^{5/2}\xi^3\frac{ i k_z }{k^2}e^{-k^2\xi^2}
\end{equation}
Constructing $\Phi_M({\bar r})$ from the inverser Fourier transform, taking  the derivative and evaluating the result at ${\bar r}=0$ gives
\begin{equation}
H^{(2)}_{eff}(r)=4M\pi^{5/2}\xi^3\int \frac{d^3k}{(2\pi)^3}\frac{k_z^2}{k^2}e^{-\frac{k^2\xi^2}{4}}=\frac{4\pi M}{3}
\end{equation}

As explained in the text the shape anisotropy term controls the domain structure but not the transition temperature.  The total field appearing in the ferromagnetic mean field equations, when represented as an exchange constant, is 
\begin{equation}
J_{FM}=(J_{SR}(Q=0)+\frac{4\pi}{3})
\label{JFM}
\end{equation}

One may compute the sums for an antiferromagnetically ordered state characterized by a wavevector $Q$ in a very similar way. For $Q\xi\gg1$ the long-ranged term vanishes  and the short ranged term is independent of $\xi$ and  we find
\begin{equation}
J_{AF}(Q)=K^q_1(Q)+K^2_2(Q)
\label{JAF}
\end{equation}
with (note the $G=0$ term is now non-vanishing)
\begin{equation}
K^q_1(Q)=-\frac{4\pi}{3V_{cell}}\sum_{G}  e^{-\frac{\left({\vec Q}+{\vec G}\right)^2}{4q^2}} \frac{3(Q_z+G_z)^2-\left({\vec Q}+{\vec G}\right)^2}{\left({\vec Q}+{\vec G}\right)^2}
\label{K1Qanswer}
\end{equation}
\begin{equation}
K^q_2(Q)=\frac{8}{3\sqrt{\pi}}\sum_{R\neq 0}\int_q^\infty d\kappa \kappa^4 e^{i{\vec Q}\cdot{\vec R}-\kappa^2 R^2}  \left(3Z^2-R^2\right)
\label{K2Qanswer}
\end{equation}

Provided that the limit is taken with  $Q\xi \gg 1$, $J_AF(Q\rightarrow 0)\rightarrow J_{FM}$. 

Finally, one may use similar methods to  compute the orientation field, Eq \ref{Horientation}. Here the short ranged terms vanish by symmetry so we are left with a term analogous to Eq \ref{Heff1} but proportional to the $x$ component of the magnetization, $M_x$. The resulting expression  may be analysed along the lines of Eq \ref{integraleqn}. The local term vanishes by symmetry and the result is 

\begin{equation}
H_{orientation}(r)=\partial_z \int d^3r^{'} \frac{1}{|r-r^{'}|}\partial_{x^{'}}M_x(r^{'})
\label{Heff2}
\end{equation}
which is just the $z$ component of the demagnetization field associated with the polarization induced the x direction by the applied transverse field. This demagnetization field would vanish for an ellipsoidal sample if $x$ is a symmetry axis and would in general be small.

\end{appendix}

\end{document}